\documentclass[12pt]{article}

\usepackage{graphicx}
\usepackage{amsmath}

\newcommand{\be}{\begin{equation}}
\newcommand{\ee}{\end{equation}}
\newcommand{\bea}{\begin{equation}\begin{aligned}}
\newcommand{\eea}{\end{aligned}\end{equation}}

\newcounter{mysect}
\newcommand{\newsect}{\par\vspace{0.5 cm}\addtocounter{mysect}{1} {\bf \arabic{mysect}.} }

\title{Hadron resonance gas and nonperturbative\\ QCD vacuum at finite temperature.}
\author{
N.O.~Agasian\thanks{agasian@heron.itep.ru} ~and S.M.~Fedorov\thanks{fedorov@heron.itep.ru} }
\date{}

\begin{document}

\maketitle
\vspace{-1cm}
\begin{center}
{\it Institute of Theoretical and Experimental Physics, \\
117218, Moscow, Russia }
\end{center}
\vspace{1cm}

\begin{abstract}
We study the nonperturbative QCD vacuum with two light quarks at
finite temperature in the framework of hadron resonance gas.
Temperature dependence of the quark and gluon condensates in the
confined phase are obtained. We demonstrate that the quark
condensate and one half (chromo-electric component) of gluon
condensate evaporate at the same temperature, which corresponds to
the temperature of quark-hadron phase transition. Critical
temperature is $T_c\simeq 190$~MeV when temperature shift of
hadron masses is taken into account.
\end{abstract}

\vspace{0.3cm}

PACS:11.10.Wx,12.38.Mh

\vspace{0.3cm}

\newsect
As well known, QCD at finite temperature undergoes a phase
transition from hadron phase, characterized by confinement and
chiral symmetry breaking, to the phase of hot quark-gluon matter.
At the critical point $T_c$ where the phase transition occurs the
behavior of the thermodynamic properties of the system, such as
energy density $\varepsilon$, specific heat, non-ideality
$(\varepsilon -3P)/T^4$, etc., is drastically changed.
More than that, the phase transition in QCD
is characterized by the radical rearrangement of the
non-perturbative quark-gluon vacuum.

Lattice calculations for finite temperature QCD show that the
deconfinement and chiral invariance restoration take place at the
same temperature, and for the case of two light quarks ($N_f=2$)
critical temperature is in the interval $T_c \sim 175 \div
190$~MeV~\cite{Karsch:2000kv,Nakamura:2003rm}. From the studies of
QCD on the lattice and from experimental data on high energy
collisions it also follows that energy density of the system at
quark-hadron phase transition is of the order of $\varepsilon_c
\sim 1 \div 1.5$~GeV/fm$^3$.

Recently it was demonstrated in the lattice simulations that for
gauge $SU(3)$ theory without quarks and for QCD with $N_f=2$
electric component of gluon condensate is strongly suppressed
above critical temperature $T_c$, while magnetic component even
slightly grows with temperature~\cite{D'Elia:2002ck}.
Also, gluon condensate at finite temperature in $SU(2)$ lattice gauge theory
was studied in the earlier paper~\cite{Lee:1989qj}. It was shown there
that approximately one half of the gluon condensate does not vanish above $T_c$.

These results are in line with theoretical predictions of the
deconfining phase transition within the
''evaporation model''~\cite{Simonov:bc} approach. Within the framework of the
effective dilaton Lagrangian at finite temperature the temperature
dependence of the gluon condensate and it's discontinuity at $T=T_c$ in
pure-glue QCD was studied in~\cite{Agasian:fn}.
Later in the paper~\cite{Agasian:2003}
temperature dependence of the gauge invariant bilocal correlator of
chromo-magnetic fields and spatial string tension
$\sigma_s(T)$ were found analytically. It was obtained that the
chromo-magnetic condensate at $T<2 T_c$ grows slowly as temperature increases,
$\langle H^2 \rangle_T = \langle H^2 \rangle_0 \coth \left( {M}/{2T} \right)$,
where $M = 1/\xi_m$ is the inverse magnetic correlation length,
which does not depend on temperature at $T<1.5 T_c$~\cite{D'Elia:2002ck}.
In the region $T>2 T_c$ magnetic correlator amplitude grows,
$\langle H^2 \rangle_T \propto g^8(T) T^4$, and correlation length decreases,
$\xi_m(T) \propto 1/(g^2(T) T)$, with increasing temperature. This behavior of
magnetic correlator~\cite{Agasian:2003} explains magnetic confinement in the framework of the stochastic
vacuum model. Obtained temperature dependence
of the spatial string tension is in perfect agreement with lattice data~\cite{Bali:1993tz}
in all temperature regions.

Recently the connection between deconfining and chiral phase transitions in QCD
was studied in~\cite{Mocsy:2003qw} in the framework of effective models which take
into account Polyakov loops dynamics. Authors discuss that phase transitions take
place at the same temperature.
Also the deconfining transition in effective model of the Yang-Mills theory with
non-order parameter fields was considered in~\cite{Sannino:2002wb}.

Thus, taking into account above listed facts, one has to obtain in
the framework of single approach that in QCD with $N_f=2$ at the critical point
$T_c \sim 175 \div 190$~MeV energy density reaches value
$\varepsilon_c \sim 1\div1.5$~GeV/fm$^3$, quark condensate $\langle \bar q q \rangle_T$
vanishes, and only one half of gluon condensate (chromo-electric component, responsible
for the formation of string and confinement) ''evaporates'', which is required to
retain magnetic confinement.

In this paper we study temperature properties of quark and gluon condensates
in the approach, based on description of the confined phase as hadron resonance
gas\footnote{Hadron resonance gas model was proposed by R.~Hagedorn~\cite{Hagedorn} for the
description of hot strongly-interacting matter.}.
We demonstrate, that all above listed phenomena can be quantitatively explained in
the framework of this approach if one takes into account temperature shift of hadron
masses.

\newsect
We will consider QCD with two light quarks. Then, knowing pressure in the
hadronic phase, $P_h(T)$, and making use of Gell-Mann-Oakes-Renner (GOR) relation,
one can find the temperature dependence of quark condensate
\be
\label{eq_qbarq_T}
\frac{\langle \bar q q \rangle_T}{\langle \bar q q \rangle_0} =
  1 - \frac{1}{F_{\pi}^2}\frac{\partial P_h (T)}{\partial m_{\pi}^2},
\ee
where $F_{\pi} = 93$~MeV is the axial $\pi$-meson decay constant.
Expression for gluon condensate
$\langle G^2\rangle_T\equiv \langle (gG^a_{\mu\nu})^2\rangle_T$
was derived in~\cite{Agasian:2001bj} starting from the renormalization group consideration
of anomalous contribution to the energy-momentum tensor in QCD with $N_f=2$ at
finite temperature. Relation connecting gluon condensate with thermodynamical pressure
in QCD is given by~\cite{Agasian:2001bj}
\be
\label{eq_G2_T}
\langle G^2 \rangle_T = \langle G^2 \rangle_0 + \frac{32 \pi^2}{b}
  \left(4-T\frac{\partial}{\partial T} -
  m_{\pi}^2 \frac{\partial}{\partial m_{\pi}^2} \right)P_h (T),
\ee
where $b=11 N_c/3 - 2 N_f/3 = 29/3$.
QCD low energy theorems~\cite{Novikov} and GOR relation, which relates mass of light quark
to the $\pi$-meson mass, were used to derive~(\ref{eq_G2_T}).
Connection between trace anomaly and thermodynamic pressure in the chiral limit and
in pure-glue QCD was also considered in~\cite{Leutwyler:cd} and~\cite{Ellis:1998kj} correspondingly.
Expressions for
$\langle \bar q q \rangle_T$ and $\langle G^2 \rangle_T$ in QCD with $N_f=3$ were obtained
in~\cite{Agasian:2002uq}. Thus, knowing pressure $P_h(T)$ as a function of temperature
and $\pi$-meson mass one can find temperature dependence of quark and gluon condensates
in the hadronic phase.

In the framework of the Chiral Perturbation Theory (ChPT) the temperature
dependence of the quark condensate was first studied in Refs.~\cite{Gasser87} and at the
three-loop level of ChPT in~\cite{Leutw88}. Using the virial expansion in a gas made of
pions, kaons and etas the temperature dependence of the quark condensate in ChPT in
two and three flavor cases was studied in~\cite{Pelaez:2002xf}.

To describe thermodynamics of QCD in the confined phase we make use of hadron
resonance gas model. In this approach all thermodynamic properties of the system
are determined by the total pressure of relativistic Bose and Fermi gases, which
describe temperature excitations of massive hadrons. The motivation of using this
method is that it incorporates all essential degrees of freedom of strongly
interacting matter. Moreover, the use of hadron resonances spectrum effectively
takes into account interactions between stable particles. Description of multiple
particle production in heavy ions collisions in the framework of hadron resonance
gas~\cite{redlich} leads to good agreement with experimental data.

Thus, pressure in the confined phase is given by
\bea
\label{eq_p_hadr_T}
{P_h} &= T \sum_i g_i \eta_i \int\frac{d^3p}{(2\pi)^3}
\ln\left( 1 + \eta_i e^{- \omega_i/T}\right),\\
\omega_i &= \sqrt{p^2+m_i^2},\\
\eta_i &= \left\{
  \begin{array}{ll}
  +1 ,& \mbox{fermions}\\
  -1 ,& \mbox{bosons}
  \end{array}
\right.
\eea
where $g_i$ is the spin-isospin degeneracy factor (e.g. $g_{\pi} = 3$, $g_N = 8$, ...).
The energy density $\varepsilon_h=T\partial P_h/\partial T-P_h$ in the hadronic
phase is given by
\be
\label{eq_eps_T}
\varepsilon_h = \sum_i g_i \int\frac{d^3p}{(2\pi)^3} \frac{\omega_i}{\exp(\omega_i/T) + \eta_i}
\ee

\newsect
To study condensates in the confined phase quantitatively the knowledge about
pressure $P_h$ dependence on light quark mass, or which is the same, on $\pi$-meson
mass is needed. In the framework of hadron resonance gas model it is
equivalent to the knowledge of masses of all resonances as functions of
pion mass. This dependence was studied numerically on the lattice, and
in the paper~\cite{Karsch:2003vd} five parameters formula, inspired by
bag model, was suggested. At the certain choice of parameters it
accurately describes masses of all considered by the authors~\cite{Karsch:2003vd}
particles
\bea
\label{m_i_T}
&m_i = N_u a_1 x \sqrt{\sigma} + \frac{m_{h}}{1 + a_2 x + a_3 x^2 + a_4 x^3 + a_5 x^4},\\
&x \equiv m_{\pi}/\sqrt{\sigma},\\
&a_1 = 0.51,\quad  a_2 = a_1 N_u \sqrt{\sigma}/m_h,\\
&a_3 = 0.115,\quad a_4 = -0.0223,\quad a_5 = 0.0028.
\eea
Here $m_h$ is the physical hadron mass, $N_u$ is the number of light quarks
($N_u=2$ for mesons, $N_u=3$ for baryons), $\sigma=(0.42~\mbox{GeV})^2$ is the string tension.

Next, it should be taken into account that as temperature increases hadron masses change.
In the framework of finite temperature conformally generalized nonlinear sigma model with light
and massive hadrons~\cite{Agasian:1997zr} it was shown, that temperature shift of hadron masses
can be taken into account by the following substitution
\bea
\label{chi}
&m_h \to m_h (\chi_T/\chi_0), \quad m_{\pi} \to m_{\pi}\sqrt{\chi_T/\chi_0}, \\
&\chi_T/\chi_0=\left(\langle G^2 \rangle_T/\langle G^2 \rangle_0\right)^{1/4},
\eea
where $\chi$ is the dilaton field. Different as compared to other particles dependence
of $\pi$-meson mass is the reflection of it's goldstone nature. In the chiral limit,
$m_q \to 0$, presented relation for the hadron masses is a strict consequence of
low-energy QCD theorems. Similar to~(\ref{chi}) relations for the shift of nucleon
mass were obtained using effective dilaton Lagrangian at finite baryon
density~\cite{Agasian:1999id}.

\newsect
Formulas~(\ref{eq_qbarq_T})-(\ref{chi}) define thermodynamic properties of
the system in hadronic phase and allow to calculate quark and gluon condensates
in the whole region of temperatures below $T_c$.

We take into account all hadron states with masses below $2.5$~GeV for mesons
and $3.0$~GeV for baryons. Altogether it amounts to 2078 states (with degeneracy factors
$g_i$ taken into account). It is clear, that at low temperatures $T<m_{\pi}=140$~MeV main
contribution to the thermodynamic quantities will come from thermal excitations of $\pi$-mesons,
since other states are substantially heavier and are exponentially suppressed with Boltsman factor
$\propto \exp \{-m_h/T\}$. However, a great many of heavy states starts playing important role
when $T>m_{\pi}$. In figure~\ref{fig_Ph} pion contribution to the pressure is shown with dash-dotted line. It is seen, that
below temperature $T=120$~MeV pions give main contribution to the $P_h$. At higher temperatures main
contribution comes from all other hadron states. Figure~\ref{fig_Ph} also shows lattice data~\cite{Karsch:2000ps}
for the pressure $P_h$ in QCD with $N_f=2$. One can see that in the region $T<T_c$ hadron resonance
gas model with the account for temperature mass shift gives good description of pressure as
a function of temperature.

In figure~\ref{fig_epsh} energy density $\varepsilon_h$ as a function of temperature is presented.
Value of 1~GeV/fm$^3$, corresponding to the estimates for energy density at quark-hadron phase
transition, is reached at temperature $T\simeq 175$~MeV, i.e. in the region of phase transition
temperature, as obtained in lattice calculations~\cite{Karsch:2001cy}.

Figures~\ref{fig_qq} and~\ref{fig_G2} show quark and gluon condensates as  functions of temperature.
Shaded area in figure~\ref{fig_G2} corresponds to zero-temperature values of gluon condensate
being in the range $\langle G^2\rangle_0 = (0.5 \div 1.0)$~GeV$^4$.
It is important that quark condensate goes to zero at the same temperature, where half of
gluon condensate evaporates, and if temperature shift of hadron masses is taken into account,
this temperature is $T\simeq 190$~MeV.

\unitlength=1cm
\begin{figure}[h]
\begin{picture}(11,6.5)
\put(0.7,0.5){\includegraphics[width=9cm]{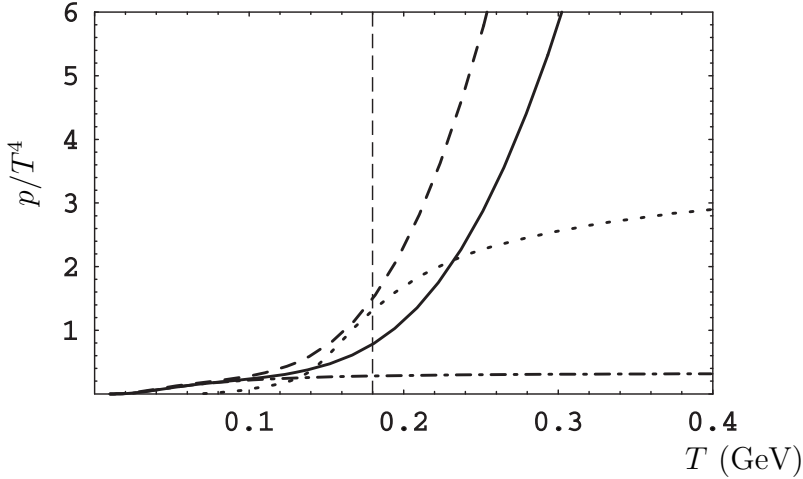}}
\put(9,0){$T$ (GeV)}
\put(0,3.5){\rotatebox{90}{$p/T^4$}}
\end{picture}
\caption{Pressure $P_h/T^4$ as a function of temperature. Solid line -- zero temperature hadron spectrum;
dashed line -- spectrum with temperature shift taken into account, $\chi_T/\chi_0 = 0.84$;
dash-dotted line -- pion excitations only. Dotted line -- lattice data~\cite{Karsch:2000ps}.}
\label{fig_Ph}
\end{figure}

\begin{figure}[h]
\begin{picture}(11,6.0)
\put(0.7,0.5){\includegraphics[width=9cm]{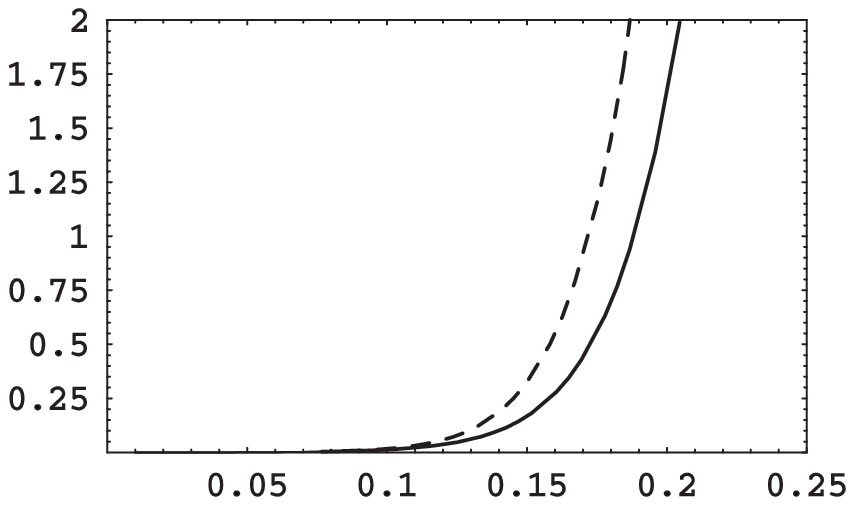}}
\put(9,0){$T$ (GeV)}
\put(0,2.5){\rotatebox{90}{$\varepsilon_h$ (GeV/fm$^3$)}}
\end{picture}
\caption{Energy density $\varepsilon_h$ as a function of temperature.
Solid line -- zero temperature hadron spectrum;
dashed line -- spectrum with temperature shift taken into account, $\chi_T/\chi_0 = 0.84$.}
\label{fig_epsh}
\end{figure}

\begin{figure}[h]
\begin{picture}(11,6.5)
\put(0.7,0.5){\includegraphics[width=9cm]{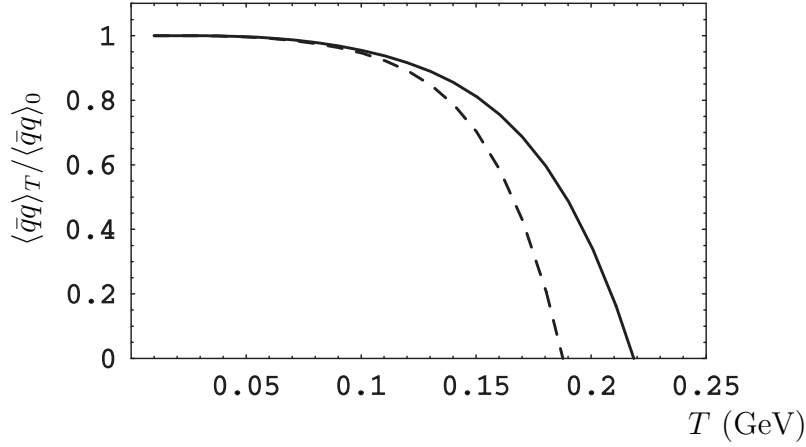}}
\put(0,2.5){\rotatebox{90}{$\langle \bar q q \rangle_T/\langle \bar q q \rangle_0$}}
\put(9,0){$T$ (GeV)}
\end{picture}
\caption{Quark condensate $\langle \bar q q \rangle_T/\langle \bar q q \rangle_0$ as a function of temperature.
Solid line -- zero temperature hadron spectrum;
dashed line -- spectrum with temperature shift taken into account, $\chi_T/\chi_0 = 0.84$.}
\label{fig_qq}
\end{figure}

\begin{figure}[h]
\begin{picture}(11,6.5)
\put(0.7,0.5){\includegraphics[width=9cm]{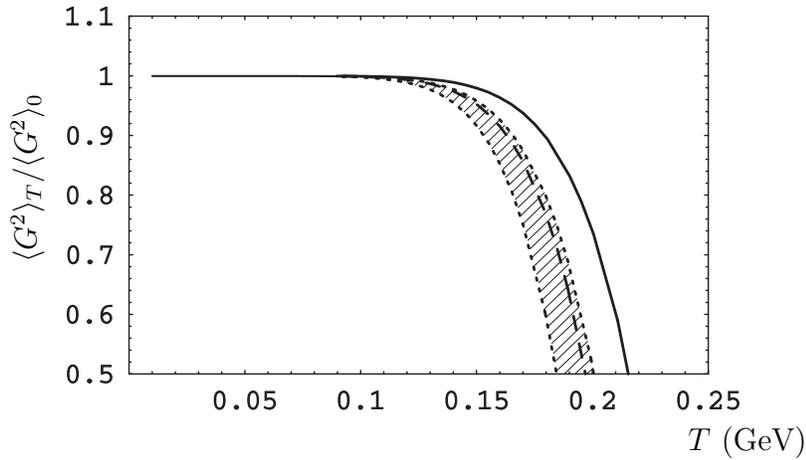}}
\put(0,2.5){\rotatebox{90}{$\langle G^2 \rangle_T/\langle G^2 \rangle_0$}}
\put(9,0){$T$ (GeV)}
\end{picture}
\caption{Gluon condensate $\langle G^2 \rangle_T/\langle G^2 \rangle_0$ as a function of temperature.
Solid line -- zero temperature hadron spectrum;
dashed line -- spectrum with temperature shift taken into account, $\chi_T/\chi_0 = 0.84$;
$\langle G^2\rangle_0 = 0.87$~GeV$^4$~\cite{Narison}.
Shaded area -- uncertainty due to zero-temperature gluon condensate value
($\langle G^2 \rangle_0 = 0.5 \div 1.0$~GeV$^4$).}
\label{fig_G2}
\end{figure}

\clearpage

Strictly speaking, one has to find change of gluon condensate in a self-consistent way
with the use of effective dilaton lagrangian at $T\neq 0$ and taking into account shift
of hadron masses (see~\cite{Agasian:1997zr}). However, numerical calculations show that
up to temperatures $T \sim m_{\pi}$ gluon condensate decreases very slowly, and at
$T=m_{\pi}$, $\Delta \langle G^2 \rangle_T \approx 0.02 \langle G^2 \rangle_0$. As temperature
grows, gluon condensates drops abruptly and changes by $\sim 50 \% $ in a small temperature
region $\Delta T \sim 50$~MeV. Accordingly, we present calculations with temperature
shift of hadron masses being $16\%$ ($\chi_T/\chi_0 = 0.84 \simeq (0.5)^{1/4}$). Note, that
even if temperature shift of $m_h$ is not taken into account, quark condensate
and half of gluon condensate evaporate at the same temperature $T \sim 215$~MeV.

\newsect
In the present paper we have studied nonperturbative QCD vacuum with two light quarks at
finite temperature in the framework of hadron resonance gas model. We have found temperature
dependencies of quark and gluon condensates in the confined phase, and it was shown that
quark condensate and one half of gluon condensate (chromo-electric component) evaporate
at the same temperature, which corresponds to quark-hadron phase transition. This fact
confirms picture of magnetic confinement, i.e. that chromo-electric condensate vanishes, while
chromo-magnetic condensates almost does not change at the phase
transition~\cite{D'Elia:2002ck,Simonov:bc,Agasian:fn,Agasian:2003}. The energy density of hadron
resonance gas at transition temperature is $\varepsilon_h(T_c) \sim 1.5$~GeV/fm$^3$. When
temperature shift of hadrons masses is taken into account critical temperature is
$T_c\simeq 190$~MeV.

\par
This work is supported by NSh-1774.2003.2  grant and by
Federal Program of the Russian Ministry of Industry, Science
and Technology No 40.052.1.1.1112.

\par

\end{document}